# Digitization of a random signal from the interference of laser pulses

Issue of randomness extraction for a quantum random number generator

Roman Shakhovoy
Faculty of Networks and Communication Systems
MTUCI
Moscow, Russia
also with
NTI Center for Quantum Communications
NUST MISIS
Moscow, Russia
r.shakhovoy@goqrate.com

*Abstract*—In the study of quantum random number generators (QRNGs), the problem of random signal digitization is often not considered in detail. However, in the context of a standalone QRNG device, this issue is very important. In this paper, we consider the problem of digitizing laser pulses with random intensity and analyze various approaches used to estimate the contribution of classical noise. A simple method for determining the quantum reduction factor suitable for digitization with an analog-to-digital converter is proposed.

## I. Introduction

A special place among physical (hardware) random number generators (RNGs) is occupied by quantum RNGs (QRNGs) based on the digitization of noise from a quantum entropy source. Over the past two decades, a number of different QRNGs based on various quantum effects have been demonstrated [1]. In this list, the most convenient and relatively inexpensive QRNGs are those that get quantum randomness from laser radiation. Here, the source of entropy is spontaneous emission, whose properties are determined by their relation to zero-point (vacuum) fluctuations of the electromagnetic field [2, 3], which are generally considered to be perfectly uncorrelated and broadband.

The vast majority of QRNGs of this type converts phase fluctuations induced by the spontaneous noise into random changes in light intensity, which can be readily measured with classical photodetectors. One of the convenient methods for obtaining a random signal in this kind of QRNGs is the interference of laser pulses emitted by a gain-switched semiconductor laser [4-7]. Under gain switching, each new laser pulse appears with a random phase, so the interference of pulses coming out of the laser at different times will be completely random. From a physical point of view, the random bit generation rate in this case is limited by the repetition rate of laser pulses. It was pointed out in [8] that, with a suitable choice of the pump current, the pulse repetition rate can be increased up to 10 GHz in such QRNGs. However, from a technical point of view, a high pulse generation rate imposes high requirements on the photodetector and digitizer, whose bandwidth must be sufficient for processing such signals. While photodetectors of 20-30 GHz bandwidth are not very expensive and are available on the market, the bandwidth of modern analog-to-digital converters (ADCs) is limited to several gigahertz, and devices with an analog bandwidth of more than 2 GHz are quite expensive and not always available on the market. Therefore, when creating a real QRNG, one should consider the effects associated with the finite bandwidth of digitizing devices.

In this paper, I study the effect of an ADC with the finite bandwidth on the statistical properties of the digitized signal corresponding to the random pulse interference. It is shown that the effect of a finite bandwidth is significantly enhanced in the presence of jitter, especially when short laser pulses are used. In addition, the problem of quantum noise extraction and the dependence of the quantum reduction factor on the ADC bit depth are considered. To the best of the author's knowledge, these issues have not been previously studied in the literature.

## II. Schematic of the Interference-Based QRNG

The possible realization of the interference-based QRNG is shown in Fig. 1. It includes a gain-switched single-frequency laser diode generating optical pulses that are sent to the interferometer, whose delay line is chosen such that pulses emitted by the laser at different times are met at the interferometer's output. (In other words, the length of the delay line $\Delta L$ should provide the time delay, which is a multiple of the pulse repetition period). To improve spectral matching between interfering pulses, one may install a bandpass optical filter somewhere between the laser and the photodetector (see [9] for more details). In case of a Michelson interferometer, it is also desirable to use an optical isolator to prevent the reflected optical signals from entering the laser (it is also possible to use an optical circulator for this purpose).

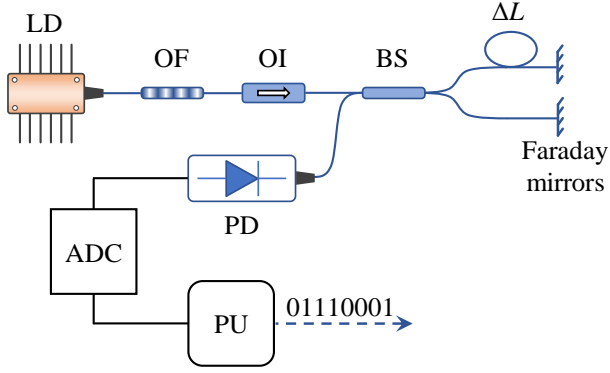

Fig. 1. Simplified schematic of the interference-based QRNG: LD – laser diode, OF – optical filter, OI –optical isolator, BS – beam splitter, PD – photodetector, ADC – analog-to-digital converter, PU – processing unit.

The result of the interference is a sequence of optical pulses of random intensity measured by the photodetector. The signal from the photodetector is fed into an ADC, where it is digitized, and then sent to the processing unit (e.g., to the field-programmable gate array), where post-processing is carried out (see Fig. 1). The processing unit accumulates the statistics of the digitized pulses, which allows determining the probability density function (PDF) of the interference signal. Using the acquired PDF, the quantum reduction factor $\Gamma_{ADC}$ is determined, which characterizes the non-uniformity of the distribution and estimates the contribution of classical noise (see next section).

## III. DIGITIZATION AND RANDOMNESS EXTRACTION

### A. Probability Density Function

Despite the simplicity of implementation, the interference of laser pulses has several features that adversely affect the visibility and can significantly affect the PDF of a random interference signal. The most important of these features are the chirp that inevitably occurs due to direct modulation of the pump current and fluctuations in the pulse emission time (jitter). Thus, it can be shown (see, e.g., [9]) that the interference of laser pulses having a Gaussian shape (and thus exhibiting linear chirp) yields (in terms of the integral signal):

$$S = s_1 + s_2 + 2\kappa_\delta \sqrt{s_1 s_2} \cos(\Delta\Phi), \quad (1)$$

where $s_1$ and $s_2$ are integral (normalized) signals corresponding to outputs from the interferometer arms, $\Delta\Phi$ is the phase difference between interfering pulses, and the coefficient

$$\kappa_\delta = \exp\left(-\frac{(1+\alpha^2)\delta^2}{8w^2}\right) \quad (2)$$

determines the interference visibility ($0 \leq \kappa_\delta \leq 1$). Here, $w$ is the Gaussian width of the laser pulse, $\alpha$ is the linewidth enhancement factor (the so-called Henry factor [10]), and $\delta$ is the inaccuracy of the pulse overlap.

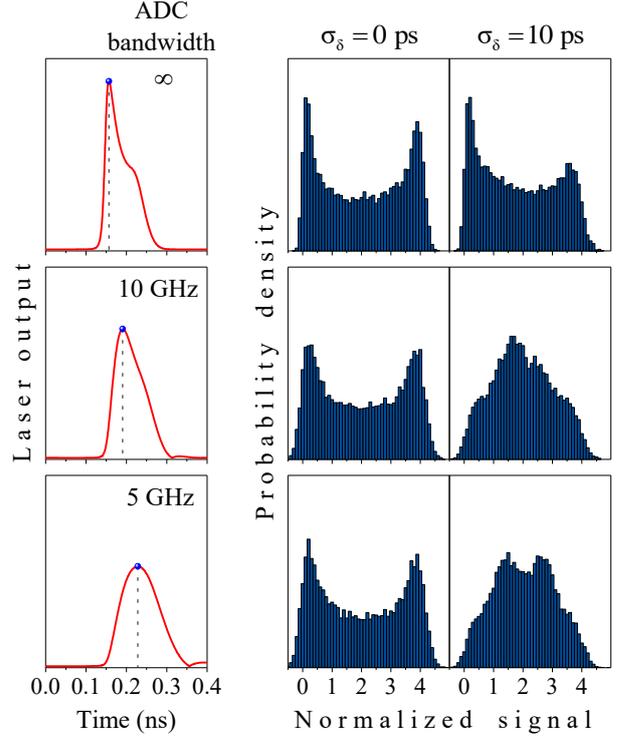

Fig. 2. Theoretical dependences of the laser pulse shape and PDF of the interference signal on the ADC bandwidth at pulse repetition rate 2.5 GHz.

It should be borne in mind that in addition to random changes in the phase difference $\Delta\Phi$, causing random changes in the interference signal, in a real experiment, it is also necessary to take into account fluctuations in the output laser power (fluctuations in the values $s_1$ and $s_2$), as well as random changes in the time of laser pulse emission (jitter). It can be seen from ☐1☐ and ☐2☐ that due to jitter (fluctuations in $\delta$), visibility becomes a random variable, so that the PDF of the integral signal $S$ will depend significantly on the PDF of jitter. The dependence on jitter will be more pronounced in the presence of a chirp, which, according to ☐2☐, increases the jitter contribution by $\sqrt{1+\alpha^2}$ times.

It is important to emphasize that relation ☐2☐ for visibility is valid only in the case of Gaussian pulses. The deviation of the pulse shape from the Gaussian curve, e.g., due to the presence of relaxation spikes, will lead to a more complex dependence of the interference signal on $\Delta\Phi$.

The cumulative distribution function (CDF) of the random signal $S$ can be defined as follows:

$$F_S(y) = \int_{S<y} f_{\Delta\Phi}(x_1) f_{s_1}(x_2) f_{s_2}(x_3) f_\delta(x_4) dx_1 dx_2 dx_3 dx_4, \quad (3)$$

to correctly display the profile of a given pulse. However, to obtain a random number, one may only know the pulse amplitude, i.e., it is enough to get a single point on its time profile. This digitization method allows the use of relatively slow (and therefore inexpensive) ADCs. Note also that multichannel ADCs can be used for digitization. In this case, different channels may be latched at different pulses, which allows the use of relatively slow ADC channels while maintaining high rates of random bit generation. Below, for simplicity, I will assume that the sampling rate of the ADC is equal to the pulse repetition rate.

From an experimental point of view, signal digitization does not imply "integration" over the pulse profile; however, due to jitter, the ADC will be triggered at slightly different moments of time for different pulses of a train. When obtaining PDF of the interference signal, this corresponds to a kind of "time integration" over some part of the pulse.

Fig. 2 shows the theoretical PDFs of a random interference signal in assumption that the laser pulse was sampled at some point on its time profile (shown by a circle on the pulse). Corresponding shapes of laser pulses are shown to the left of the PDFs. Two different time jitter values (0 and 10 ps) and three different ADC bandwidths were used in simulations. The pulse repetition rate was assumed to be 2.5 GHz. To simulate the ADC bandwidth, a second-order Butterworth filter was used. It was assumed that the bandwidth of the photodetector is significantly higher than the bandwidth of the ADC (for simplicity, the former was assumed to be infinite).

One can see from Fig. 2 that the PDF remains almost unchanged for different bandwidths in the absence of jitter. However, it depends significantly on the bandwidth of the ADC in the presence of jitter. Deviation of the PDF from the "good" one (the "good" PDF should have two well-defined maxima as shown in Fig. 2 for the case without jitter) means that the randomness caused by fluctuations in the phase of the electromagnetic field is "mixed" with additional randomness associated with jitter. The latter is generally considered to be classic in nature; therefore, quantum randomness turns out to be "contaminated" by classical fluctuations.

Let me emphasize once again that the effect of jitter is enhanced by the chirp. As shown in [9], the combined effect of chirp and jitter on the PDF of the interference signal can be reduced by filtering the high-frequency part of the spectrum. (For this purpose, the optical schematic in Fig. 1 contains the bandpass optical filter.) It is important to note that with a decrease in the pulse repetition rate, the effect of jitter "contamination" of quantum noise can be avoided even without using a bandpass filter if longer laser pulses are generated. Thus, if the repetition period of laser pulses is 2 ns (pulse repetition rate of 500 MHz), then it will not be difficult to generate laser pulses with a width of the order of 1 ns. Since in most conventional DFB lasers transients end at times of the order of 200–250 ps after the onset of lasing, most of the pulse will not be chirped and, therefore, jitter will not have a significant effect on the interference.

Fig. 3 shows the dependences of the PDF similar to those shown in Fig. 2, but for much longer laser pulses, assuming a repetition rate of 500 MHz. It is clear from Fig. 3 that jitter

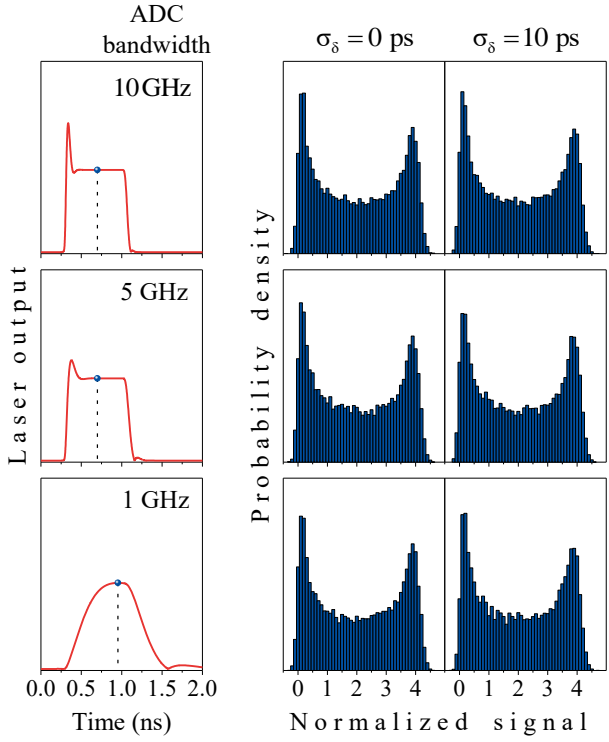

Fig. 3. Theoretical dependences of the laser pulse shape and PDF of the interference signal on the ADC bandwidth at pulse repetition rate 500 MHz.

where values of random variables $\Delta\Phi$, $s_1$, $s_2$, and $\delta$ are denoted by $x_1$, $x_2$, $x_3$, and $x_4$, respectively, and where it is assumed that fluctuations of these variables are independent, so that the resulting PDF is the product of the corresponding PDFs: $f_{\Delta\Phi} f_{s_1} f_{s_2} f_\delta$. The region of integration is determined by the inequality $x_2 + x_3 + 2\exp\left(-\frac{(1+\alpha^2)x_4^2}{8w^2}\right)\sqrt{x_2 x_3}\cos(x_1) < y$

(we still assume here the interference of Gaussian pulses). The resulting distribution density function of the interference signal is determined by the derivative of the distribution function: $f_S = F_S'$. Unfortunately, the integral in (3) cannot be found analytically, so Monte-Carlo simulations are generally used to find $f_S$.

In a real experiment, the PDF of the interference signal is additionally "broadened" due to the noise of the photodetector. To take this into account, the signal measured with the photodetector can be written in the following form:

$$S \to S + \zeta, \quad (4)$$

where $\zeta$ is the classical Gaussian noise of the photodetector.

### B. Digitization of a Random Signal

A researcher working with optical pulses measured by a photodetector and displayed on an oscilloscope generally assumes that the sampling rate of the digitizer is much higher than the pulse repetition rate, since several points are required does not have a significant effect on the PDF if the sampling point (the circle on the pulse profile in Fig. 3) is chosen far enough from the beginning of the pulse. Moreover, this is true even for ADCs with a relatively small (~1 GHz) analog bandwidth.

So, to obtain a random number from a laser pulse with a random intensity, it is not necessary to digitize the entire pulse, but it is enough to get a single sampling point at a properly chosen position on its time profile. This greatly simplifies the problem and, importantly, significantly reduces the requirements for the sampling rate of the ADC. Thus, to digitize a random signal one can use relatively slow (and therefore inexpensive) ADCs with a sampling rate of several hundred MSa/s. Meanwhile, as shown in Fig. 3, it is enough for such an ADC to have an analog bandwidth of just about 1 GHz.

*C. Quantum Reduction Factor*

The bit sequence obtained directly via digitization of a random physical signal is not yet truly random (for a more rigorous definition of true randomness in the context of a QRNG, see [7]). This already follows from the fact that the PDF of a random signal is usually not uniform, which entails an uneven bit sequence. A physical source of randomness with a non-uniform PDF of the measured signal is sometimes called a weak entropy source, usually meaning a cryptographic "weakness" of the secret key obtained with it. Raw random sequences obtained from weak entropy sources are unsuitable for cryptographic applications.

So, a raw random sequence must be subjected to mathematical processing, after which it should become (almost) uniform. The algorithms used for this are called randomness extractors. The randomness extractor can be defined as a function that takes a binary string of length $N$ as input and returns a sequence of length $M$ (by definition, the inequality $M < N$ must be satisfied), i.e., the randomness extraction procedure, "compresses" the raw sequence, making it more random. (The value $\gamma = N/M$ is sometimes referred to as the reduction factor.) More formally, a randomness extractor can be written as a function $E : \{0,1\}^N \to \{0,1\}^M$, where the notation $\{0,1\}^{N,M}$ corresponds to binary vectors of length $N$ and $M$, respectively. Obviously, the more the distribution of random bits differs from uniform and the more correlations are present in the binary sequence, the larger the reduction factor should be, and, therefore, the less randomness can be extracted. Usually, in randomness extraction procedures, the reduction factor $\gamma$ is estimated in terms of the min-entropy $H_\infty$ per element (i.e., per word) of the raw bit sequence. Thus, an ideal randomness extractor applied to a non-uniform random sequence $\{X_1, X_2, ..., X_N\}$ with $N \gg 1$, where each $X_i$ is an $n$-bit word, can provide $NH_\infty$ of almost uniformly distributed bits [11]. In other words, the raw sequence will be reduced by such an extractor by a factor of $\gamma$, where

$$\gamma = \frac{n}{H_\infty}. \quad (5)$$

The min-entropy in (5) is defined as $H_\infty = -\log_2 p_{max}$, where $p_{max}$ is the highest probability of "guessing" some random element from the sequence $\{X_1, X_2, ..., X_N\}$. If a random signal is digitized using an ADC, then $n$ in (5) may be taken to be equal to the ADC bit depth, while $p_{max}$ will be then correspond to the probability of the most probable bin (the number of bins in this case is, obviously, $2^n$).

In the case of QRNG, the problem of extracting randomness goes beyond the definition given above. Indeed, the classical procedure for extracting randomness comes down, in essence, to eliminating the uneven distribution of bits in a raw sequence. If the task is to extract quantum randomness, then one should consider not only the non-uniformity of the distribution, but also the inevitable contribution of classical noise mixed with the signal received from the source of quantum entropy.

The need for classical noise filtering is because an adversary (Eve) can, in principle, control classical noise in a system with QRNG. Therefore, she can get some information about the raw sequence of random bits. One can, e.g., imagine a situation where a quantum device is connected to an external power source whose voltage fluctuations are ~1% of the nominal value. It may turn out that the source of these fluctuations is Eve, who in this case knows exactly the true value of the voltage at each moment of time, which means that she can potentially receive some part of the information about the raw sequence. Therefore, it is necessary to somehow separate the quantum component of the signal from the classical one.

It is extremely difficult to do this directly at the physical level, since measuring devices (photodetector, ADC) are classical. Therefore, it is generally assumed that the presence of a "classical component" in a raw sequence is equivalent to the presence of hidden correlations, which can be eliminated by processing the sequence with a randomness extractor. Therefore, the reduction factor must now be chosen such that not only the non-uniformity of the raw sequence, but also the assumed hidden correlations are eliminated after the randomness extraction procedure. Obviously, this problem cannot be solved only by mathematical methods designed to work with binary sequences. Indeed, looking at random sequences of bits, it is impossible to determine which source of entropy was used to obtain them: quantum, classical, or pseudo-random. Therefore, a specific physical model of the entropy source should be included in the procedure for extracting randomness in QRNG. To distinguish the quantum reduction factor from the factor $\gamma$ defined above, we denote it by $\Gamma$. It is obvious that $\Gamma \geq \gamma$, and equality can be achieved only in the absence of classical noise.

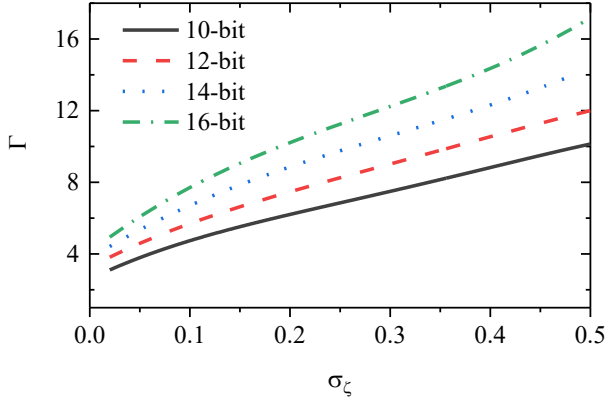

Fig. 4. Theoretical dependences of the quantum reduction factor defined by (12) on the standard deviation of the photodetector noise when using ADCs of different bit depth.

In [7], the use of the quantum reduction factor was considered in detail for the case of digitizing a random signal by a comparator. In this case, $\Gamma$ can be defined as follows:

$$\Gamma = \frac{1}{2 - H_\infty}, \quad (6)$$

where the min-entropy is defined by

$$H_\infty = -\log_2\left(\int_{S_{min}}^{S_{min}+w_{\Delta\Phi}/2} f_S(x)\,dx\right), \quad (7)$$

where, as above, $f_S$ is the experimental PDF of the interference signal, which generally depends on fluctuations in the laser output power, jitter, photodetector noise, etc. In [7], $S_{min}$ and $w_{\Delta\Phi}$ are included in the definition of the "ideal" (quantum) PDF $f_S^Q$ of the interference signal $S$ defined by (1) under the assumption that there are no other noises except fluctuations of $\Delta\Phi$. The latter is given by

$$f_S^Q(x) = \left[\pi\sqrt{(x - S_{min})(S_{max} - x)}\right]^{-1}, \quad (8)$$

where

$$S_{min} = s_1 + s_2 - 2\kappa_\delta\sqrt{s_1 s_2},$$
$$S_{max} = s_1 + s_2 + 2\kappa_\delta\sqrt{s_1 s_2}, \quad (9)$$

and $w_{\Delta\Phi} = S_{max} - S_{min}$.

In [7], the definition of the quantum reduction factor was generalized for the case of an ADC:

$$\Gamma = \frac{n}{1 + H_\infty^Q - H_\infty}, \quad (10)$$

where

$$H_\infty = -\log_2\left(\int_{S_{min}}^{S_{min}+\Delta U/2^n} f_S(x)\,dx\right), \quad (11)$$

and $H_\infty^Q$ is also given by (11) but with $f_S^Q$ under the integral. In (11), $\Delta U$ is the ADC input voltage range, and $n$ is its bit depth. The physical meaning of (10) is similar to the meaning of (6): as soon as the probability that a signal value corresponding to a bin in the vicinity of $S_{min}$ decreases by a factor of 2 due to the contribution of classical noise, we stop trusting the digitized noise. This approach, however, corresponds to an extremely tough restriction on the presence of classical noise. Thus, using (1), (10) and (11) it can be shown that even for a 10-bit ADC, the quantum reduction factor $\Gamma$, defined by (10), tends to infinity already at relative photodetector noise (in terms of standard deviation) $\sigma_\zeta \sim 0.3\%$ (such an estimate was obtained for fluctuations of laser output power of $\sigma_{s_1} = \sigma_{s_2} = 1\%$). For an ADC with a larger bit depth, the quantum reduction factor from (10) grows even faster with an increase in classical noise. Thereby, (10) can hardly be used to estimate $\Gamma$ in real systems, where the noise level is generally much higher than 0.3%.

One may formulate an alternative definition of the quantum reduction factor for an ADC:

$$\Gamma = \frac{n}{2H_\infty^Q - H_\infty}. \quad (12)$$

The physical meaning of this formula is somewhat different: we stop trusting a digitized random signal when the min-entropy $H_\infty$ corresponding to the probability that the value of the pulse intensity falls into the bin in the vicinity of $S_{min}$ doubles (i.e. when $H_\infty$ becomes equal to $2H_\infty^Q$). This condition corresponds to a less stringent constraint on the presence of classical noise.

Fig. 4 shows the theoretical dependences of the factor $\Gamma$ defined by (12) on the standard deviation of the photodetector noise $\sigma_\zeta$. In simulations, the standard deviation of the relative fluctuations of the laser output power ($\sigma_{s_1}$ and $\sigma_{s_2}$) was set to 0.05. Fig. 4 shows that such $\Gamma$ still has very large values; therefore, the advantage of the ADC (obtaining several bits when digitizing one laser pulse) is almost completely leveled at the post-processing stage.

It should be emphasized that definitions (10) and (12) do not have a rigorous mathematical justification and, in essence, are only an attempt to generalize the result written in definition (6) to the case of an ADC. Here, I propose a more rigorous approach for estimating the quantum reduction factor, where $\Gamma$ is defined as the product of several factors, each of which corresponds to the contribution of one of three effects: 1) uneven distribution of the random signal, 2) ADC internal noise, which is given by the ENOB parameter (effective number of bits), and 3) photodetector and laser noise.

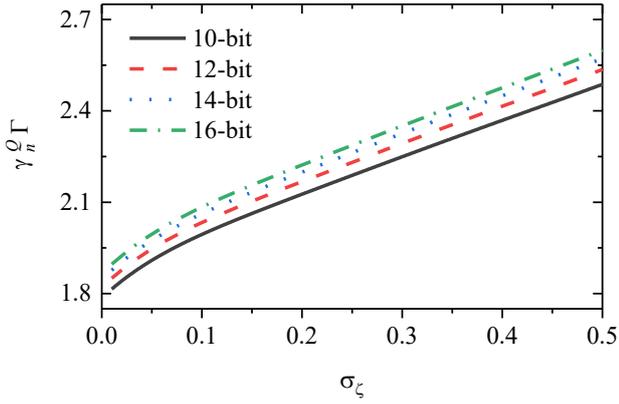

Fig. 5. Theoretical dependences of the product $\gamma_n^Q \Gamma$ defined by (6) and (13) on the standard deviation of the photodetector noise when using ADCs of different bit depth.

The intermediate reduction factor associated with the non-uniformity of the PDF will be denoted as $\gamma_n^Q$ and defined similarly to (5):

$$\gamma_n^Q = \frac{n}{H_\infty^Q} \qquad (13)$$

with $H_\infty^Q$ given by (11) but with $f_S^Q$ under the integral. Substituting (8) into (11), one can readily find:

$$H_\infty^Q = -\log_2\left\{\frac{1}{2} - \frac{1}{\pi}\arctan\left[\frac{r2^n - 2}{2\sqrt{r2^n - 2}}\right]\right\}, \qquad (14)$$

where $r = w_{\Delta\Phi}/\Delta U$.

The next reduction factor will be denoted as $\gamma_{\text{ENOB}}$ and will be defined as

$$\gamma_{\text{ENOB}} = \frac{n}{\text{ENOB}}, \qquad (15)$$

where ENOB stands for effective number of bits. The value set by the ENOB parameter corresponds to that part of the digitized signal that is guaranteed not to be "contaminated" by noise and distortions caused by imperfections in the internal components of the ADC. Thereby, $\gamma_{\text{ENOB}}$ can be considered as a reduction factor related to classical noise within the ADC. Usually, the ENOB value is given in the specifications of a particular device and depends on the ADC bit depth, as well as on the value of the SINAD (signal-to-noise-and-distortion ratio) parameter [12]:

$$\text{ENOB} = \frac{\text{SINAD} - 1.76}{6.02}. \qquad (16)$$

Finally, to determine the contribution of noise from the photodetector and laser, it seems reasonable to assume that such noise does not depend on the digitization method. One can thus use here the quantum reduction factor defined for the comparator, which is given by (6). So, the quantum reduction factor for the ADC can be written in the following form:

$$\Gamma_{\text{ADC}} = \gamma_n^Q \gamma_{\text{ENOB}} \Gamma. \qquad (17)$$

Dependences of the quantity $\Gamma_{\text{ADC}}/\gamma_{\text{ENOB}} = \gamma_n^Q \Gamma$ on the Gaussian classical noise at the photodetector are shown in Fig. 5. As above, the standard deviation of the relative fluctuations in the laser output power was set to 0.05.

It is not convenient to directly measure the contribution of classical fluctuations occurring in a laser and a photodetector, especially when it is necessary to estimate it "on the fly" (during the QRNG operation). It seems preferable to estimate this contribution comparing the PDF of a real interference signal with the ideal (quantum) PDF given by (8). It was shown in [7] that such a comparison can be carried out quite effectively by determining the ratio of the total width of the experimental PDF to the "distance" between its maxima (this parameter will be denoted as $B$). Fig. 6 shows the dependences of the value of $\Gamma_{\text{ADC}}/\gamma_{\text{ENOB}} = \gamma_n^Q \Gamma$ on the coefficient $B$ under the assumption that the standard deviation of the relative fluctuations in the output power of the laser is fixed and equal to 5%.

So, to determine the total quantum reduction factor $\Gamma_{\text{ADC}}$, one should first find the value of $B$, then determine the corresponding value of $\gamma_n^Q \Gamma$ using the plot in Fig. 6, and multiply the found value by $\gamma_{\text{ENOB}}$. The typical values of $\gamma_{\text{ENOB}}$ are usually in the range of 1.5–1.8, so the value of $\Gamma_{\text{ADC}}$ will not significantly exceed 4 even with considerable noise in the photodetector and laser.

### D. Post-processing

In cryptographic applications, the randomness extractor is generally used in the form of a family of 2-universal hash functions, whose efficiency is guaranteed by the leftover hash lemma [11]. A common way to implement 2-universal hashing is to multiply the input raw sequence by a random binary matrix [13]. Without loss of generality, one may always use for these purposes random Boolean Toeplitz matrices, which allow significantly saving the seed length. In our case, the randomness extraction is then divided into three steps:

1. For a raw binary sequence of length $n$, determine the length of the output sequence $m$ by the formula $m = n/\Gamma_{\text{ADC}}$.

2. Generate the Toeplitz matrix using a random seed of length $m+n-1$ bits.

3. Multiply the Toeplitz matrix by the raw sequence. This yields the resulting random bit string.

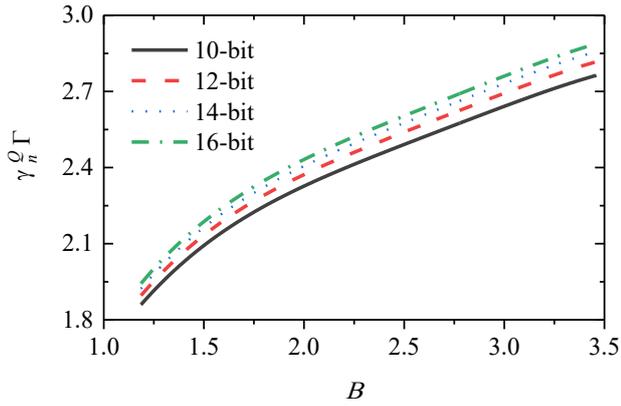

Fig. 6. Theoretical dependences of the product $\gamma_n^Q \Gamma$ defined by (6) and (13) on the ratio of the total width of the experimental PDF to the "distance" between its maxima when using ADCs of different bit depth.

To obtain the seed, one may buffer a raw random sequence of a relatively small length. Then, this sequence is subjected to a deterministic extractor, e.g., the von Neumann extractor [14]. The random sequence obtained after the extractor can be now used as a seed in hashing algorithms. "Equipped" with such a procedure, the QRNG becomes an autonomous source of entropy that does not need an additional entropy source.

## IV. CONCLUSIONS

The influence of jitter and the bandwidth of the digitizing device on the probabilistic properties of the interference of laser pulses with a random phase is studied. It is shown that when digitizing short pulses, insufficient bandwidth can significantly distort the PDF if the system is subject to jitter. An effective way to level this effect is use longer laser pulses. In addition, a simple method for estimating the contribution of classical noise based on the calculation of the quantum reduction factor is proposed. The latter is proposed to be evaluated as the product of three factors related to non-uniform distribution of the random signal, ADC internal noise, and photodetector and laser noise. The dependence of the quantity $\Gamma_{ADC}/\gamma_{ENOB} = \gamma_n^Q \Gamma$ on the ratio of the total width of the experimental PDF to the "distance" between its maxima was calculated; the corresponding plot can be used to estimate the quantum reduction factor in real devices.